\documentclass[a4paper,11pt]{article}
\usepackage{pos}
\usepackage[english]{babel}
\usepackage{bm}
\usepackage{bbm}
\usepackage{mathtools}
\usepackage{graphicx}
\usepackage{tikz}
\usetikzlibrary{arrows,shapes,backgrounds,calc,positioning,tikzmark,patterns,automata,decorations.markings}
\tikzset{fontscale/.style={font=\relsize{#1}}}
\tikzset{->-/.style={decoration={
  markings,
  mark=at position #1 with {\arrow{>}}},postaction={decorate}}}
\tikzset{-<-/.style={decoration={
  markings,
  mark=at position #1 with {\arrow{<}}},postaction={decorate}}}

\tikzset{cross/.style={cross out,draw,minimum size=2*(#1-\pgflinewidth),inner sep=0pt, outer sep=0pt}}

\tikzset{
  pics/carc/.style args={#1:#2:#3}{
    code={
      \draw[pic actions] (0,0) -- (#1:#3) arc(#1:#2:#3) -- cycle;
    }
  }
}
\usepackage{picture}
\usepackage{cancel}
\usepackage{tabularx}
\usepackage{tensor}
\usepackage{float}
\usepackage{fancyhdr}
\usepackage{sidecap}
\usepackage{xcolor}
\usepackage{scalerel}
\usepackage{relsize}
\usepackage[absolute,overlay]{textpos}
\usepackage[customcolors]{hf-tikz}
\usepackage{cals}
\usepackage{enumitem}
\setlist{nolistsep}
\usepackage{pifont}
\usepackage[export]{adjustbox} 
\usepackage[most]{tcolorbox}
\usepackage{varwidth}
\tcbuselibrary{skins}

\newtcolorbox{hlbox}[2][red]{
colbacktitle=#1!10,
colback=white!95!#1,
coltitle=black,
fonttitle=\bfseries,
colframe=#1!50,
boxrule=0.5pt,
titlerule=0pt,
title={\strut#2},
arc=3pt,
middle=0pt,
boxsep=0pt,
left skip=0pt,
right skip=0pt}

\def\lsim{\mathrel{\rlap{\lower4pt\hbox{\hskip1pt$\sim$}}
    \raise1pt\hbox{$<$}}}                
\def\gsim{\mathrel{\rlap{\lower4pt\hbox{\hskip1pt$\sim$}}
    \raise1pt\hbox{$>$}}}                

\let\originalleft\left
\let\originalright\right
\renewcommand{\left}{\mathopen{}\mathclose\bgroup\originalleft}
\renewcommand{\right}{\aftergroup\egroup\originalright}

\newcommand{\of}[1]{\left(#1\right)}

\newcommand{\cof}[1]{\left\{\right.#1\left.\right\}}

\newcommand{\avof}[1]{\left\langle #1\right\rangle}

\renewcommand*\[{\begin{equation}}
\renewcommand*\]{\end{equation}}

\let\oldstackrel\stackrel
\renewcommand*\stackrel[2]{{\scriptstyle\oldstackrel{#1}{#2}}}

\definecolor{emphcol}{rgb}{1.,0,0}
\let\oldemph\emph
\renewcommand*\emph[1]{\oldemph{\textcolor{emphcol}{#1}}}

\newcommand{\gGF}{g_{\rm GF}^2}
\newcommand{\MSb}{\overline{\textrm{MS}}}

\makeatletter
\pgfarrowsdeclare{open cap}{open cap}
{\pgfarrowsleftextend{+0pt}\pgfarrowsrightextend{+0.5\pgflinewidth}}
{
  \pgfmathsetlength{\pgfutil@tempdimb}{.5*\pgflinewidth-.5*\pgfinnerlinewidth}%
  \pgfsetlinewidth{\pgfutil@tempdimb}
  \pgfsetbuttcap
  \pgfsetdash{}{0pt}
  \pgfmathsetlength{\pgfutil@tempdima}{.5*\pgfutil@tempdimb+.5*\pgfinnerlinewidth}%
  \pgfpathmoveto{\pgfqpoint{0pt}{\pgfutil@tempdima}}
  \pgfpatharc{270}{90}{-\pgfutil@tempdima}
  \pgfusepathqstroke
}
\makeatother

\title{SU(2) gauge theory with $N_f=24$ quarks at non-zero mass}

\author[a,b]{Jarno Rantaharju}
\author[c,a,b]{Tobias Rindlisbacher}
\author[a,b]{Kari Rummukainen}
\author*[a,b]{Ahmed Salami}
\author[a,b]{Kimmo Tuominen}

\affiliation[a]{Department of Physics,
P.O. Box 64, FI-00014 University of Helsinki, Finland}

\affiliation[b]{Helsinki Institute of Physics,
P.O. Box 64, FI-00014 University of Helsinki, Finland}

\affiliation[c]{AEC, Institute for Theoretical Physics, University of Bern, Sidlerstrasse 5, CH-3012 Bern, Switzerland}

\emailAdd{jarno.rantaharju@aalto.fi}
\emailAdd{trindlis@itp.unibe.ch}
\emailAdd{kari.rummukainen@helsinki.fi}
\emailAdd{ahmed.salami@helsinki.fi}
\emailAdd{kimmo.i.tuominen@helsinki.fi}

\abstract{We study SU(2) gauge field theory
  with $N_f=24$ quarks. The theory is asymptotically non-free and, at vanishing
  quark mass, governed by a Gaussian fixed point at long distances.  On the other hand, at non-zero quark mass the quarks are expected to decouple at long distances and the system behaves like confining pure gauge SU(2) theory.  We study the mass spectrum of the theory as the quark mass is varied and obtain scaling laws for meson masses and string tension.  We also measure the evolution of the coupling constant at non-zero quark mass with gradient flow method. We observe unambiguously the decoupling of the quarks with the associated change of evolution of the coupling constant.
}

\FullConference{%
 The 38th International Symposium on Lattice Field Theory, LATTICE2021
  26th-30th July, 2021
  Zoom/Gather@Massachusetts Institute of Technology
}

\begin{document}
\maketitle

\section{Introduction}

The behaviour of SU($N$) gauge field theories as the energy scale is varied is largely dictated by their matter content.
Due to their applications in beyond Standard Model scenarios, asymptotically free theories
with an infrared fixed point \cite{Sannino:2004qp,Hill:2002ap,Dietrich:2005jn,Arbey:2015exa} have recently attracted
attention. On the lattice the properties of this type of theories have been studied for SU(2) gauge theory with matter fields
in the fundamental~\cite{Karavirta:2011zg,Leino:2017lpc,Leino:2017hgm,Leino:2018qvq,Amato:2018nvj}
or adjoint~\cite{Hietanen:2008mr,Hietanen:2009az,DelDebbio:2008zf,DelDebbio:2009fd,DelDebbio:2010hu,Bursa:2011ru,DeGrand:2011qd,Rantaharju:2015yva,DelDebbio:2015byq} representation.

Much less is known about theories which are not asymptotically free.  In this case the coupling constant does  not vanish at high energy, but typically diverges at a Landau pole.
For SU($N$) gauge theory with fundamental representation Dirac fermions this happens when the number of fermions $N_f$ is larger than $11N/2$.  While these theories are not directly relevant for the Standard Model, they pose a challenge for our understanding of the gauge field dynamics and the applicability of lattice computation methods.


In this work we study SU(2) gauge field theory with $N_f=24$ massive quarks.  In an earlier work we studied the evolution of the coupling at vanishing quark mass \cite{Leino:2019qwk}, with results which agreed with expectations: the coupling vanished at long distances (Gaussian IRFP), and the short distance behaviour was compatible with the Landau pole.  However, a non-vanishing quark mass introduces an additional scale to the system: while the UV properties remain to a large extent unaffected, the IR physics changes dramatically: quarks are expected to decouple at energy scales less than the quark mass, and the behaviour of the theory approaches that of the confining pure gauge SU(2) theory.  The coupling now grows in the infrared instead of vanishing.  This behaviour will have implications for the particle spectrum of the theory as the quark mass is varied.

Here we report on a study of the excitation spectrum of the theory as functions of the quark mass, and the evolution of the coupling constant at non-vanishing quark mass.  We derive a scaling law, relating the hadron and glueball masses, as well as the string tension, to the quark mass, and obtain unambiguous evidence of the decoupling of the quarks and the reversal of the coupling constant evolution.
These results have been reported in refs.~\cite{Rantaharju:2021iro,Rindlisbacher:2021hhh}.

\section{Lattice formulation}\label{sec:lattice}

We use Wilson-clover lattice action with hypercubically truncated stout smearing (HEX smearing)
\cite{Capitani:2006ni}.  The Sheikholeslami-Wohlert clover coefficient is $c_{SW} = 1$, as is often used for HEX smeared fermions \cite{Rantaharju:2015yva}. Simulations are carried out using a hybrid Monte Carlo (HMC) algorithm with leapfrog integrator and chronological initial values for the fermion matrix inversion~\cite{Brower:1995vx}.
The HMC trajectories have unit-length and the number of leapfrog steps is tuned to yield acceptance rates above 80\%.

The bare lattice gauge coupling is parametrized with $\beta_L= 4/g_{0,\text{lat}}^2$, and we use values $\beta_L\in\cof{0.25,0.001,-0.25}$.  Because Wilson fermions induce a positive shift in effective $\beta_L$ \cite{Hasenfratz:1993az,Blum:1994xb},
very small and even negative values of $\beta_L$ are needed to compensate for this effect with large number of fermions.
Lattice sizes are
$V=N_{s}^3\times N_{t}$, where $N_{s}$ and $N_{t}$ refer to the number of lattice sites
in spatial and temporal direction. These cover values $N_{t}\in\cof{32,40,48}$ and
$N_{s}\in\cof{N_{t}/2,3\,N_{t}/4,N_{t}}$.
The physical quark mass $m_q$ is measured using the lattice PCAC relation \cite{Luscher:1996ug}.

The lattice gauge coupling is defined using the gradient flow with the
``continuous $\beta$-function'' approach \cite{Peterson:2021lvb}, where the gradient flow of the gauge action determines the scale where the coupling is evaluated.  This is in contrast with the standard step scaling, where the scale is set by the lattice size \cite{Luscher:1992an}.
On a lattice of size $L^4$ the coupling is defined as a function of the flow length scale $\lambda_L=\sqrt{8\,\tau}$, where $\tau$ is the flow time, by \cite{Fodor:2012td}:
\[
\gGF\of{\lambda_L,L}=\frac{2\,\pi^2\,\lambda_L^4\,\avof{E\of{\lambda_L}}}{3\,\of{N^2-1}\of{1+\delta\of{\lambda_{L}/L}}}\ .\label{eq:latgsq}
\]
Here $\avof{E\of{\lambda_L}}$ is the expectation value of the (clover) energy of the gradient flow evolved gauge field at flow scale $\lambda_L$, and $\delta\of{c}$
is a finite volume correction. The flow is governed by the Lüscher-Weisz action~\cite{Luscher:1984xn}.
The normalization in Eq.~\eqref{eq:latgsq} is such that $\gGF$ matches the $\MSb$ running coupling at one-loop level. In commonly used renormalization schemes the first two loops are universal (including the gradient flow scheme on a lattice with Dirichlet boundary conditions \cite{Luscher:2010iy}).  However, on a periodic lattice gauge field zero modes render the two-loop term non-universal \cite{Fodor:2012td}.

\section{Mass spectrum}\label{sec:mass}

We can estimate the expected behaviour of the hadron and confinement scales as the quark mass $m_q$ vanishes by considering the solution of the 1-loop $\beta$-function:
\begin{equation}
  g^2(\mu,N_f) = \frac{1}{2\beta_0^{(N_f)} \log(\mu/\Lambda) }\,.
  \label{g1loop}
\end{equation}
Let us take $N_f=24$.  Now
$\Lambda = \Lambda_{\text{UV}}$ approximates the UV Landau pole, $\mu < \Lambda_{\text{UV}}$ and $\beta_0^{(24)} \approx -0.0549$.  Thus, when the energy scale $\mu\rightarrow 0$
the coupling constant $g^2$ vanishes, and the system is free in the infrared.

If the quarks are massive the situation changes: when $\mu \ll m_q$, the quarks decouple and the system effectively becomes confining pure gauge SU(2) theory.  The string tension is non-vanishing, and the mass spectrum includes glueballs, quark-antiquark mesons and two-quark baryons.

The energy scale of the confinement of SU(2) gauge theory sets the mass scale of glueballs and string tension.  We can estimate the confinement scale with the 1-loop
running of the coupling in $N_f=24$ and $N_f=0$ theories, and setting the couplings equal at $\mu=m_q$:
\begin{equation}
  g^2(\mu=m_q,N_f=24) = g^2(\mu=m_q,N_f=0)\,.
\end{equation}
Let us call the $\Lambda$-parameter of the $N_f=0$ theory $\Lambda_\text{IR}$.  It is analogous to ``$\Lambda_{\text{QCD}}$'' of the pure gauge theory and is a proxy for the confinement energy scale. Solving for $\Lambda_\text{IR}$ in terms of the quark mass and the UV scale $\Lambda_{\text{UV}}$ we obtain
\begin{equation}
  \frac{\Lambda_{\text{IR}}} {\Lambda_{\text{UV}}} = \left(\frac{m_q}{\Lambda_{\text{UV}}}\right)
                ^{1 - \beta_0^{(24)}/\beta_0^{(0)}}
  \approx \left(\frac{m_q}{\Lambda_\text{UV}}\right)
                ^{2.18}\,.
\label{approxLambda0}
\end{equation}
Thus, the confinement scale, and hence the glueball masses and the square root of the string tension, are proportional to $m_q^{2.18}$ at small quark masses.
While the approximation (\ref{approxLambda0}) is based on 1-loop running and an abrupt mass threshold, it becomes exact in the limit $m_q/\Lambda_\text{UV} \rightarrow 0$ because the coupling near $\mu=m_q$ will be small, and the small coupling region dominates the evolution of the scale.

\begin{figure}[tbp]
\centering
\hfill
\includegraphics[height=0.35\linewidth,keepaspectratio]{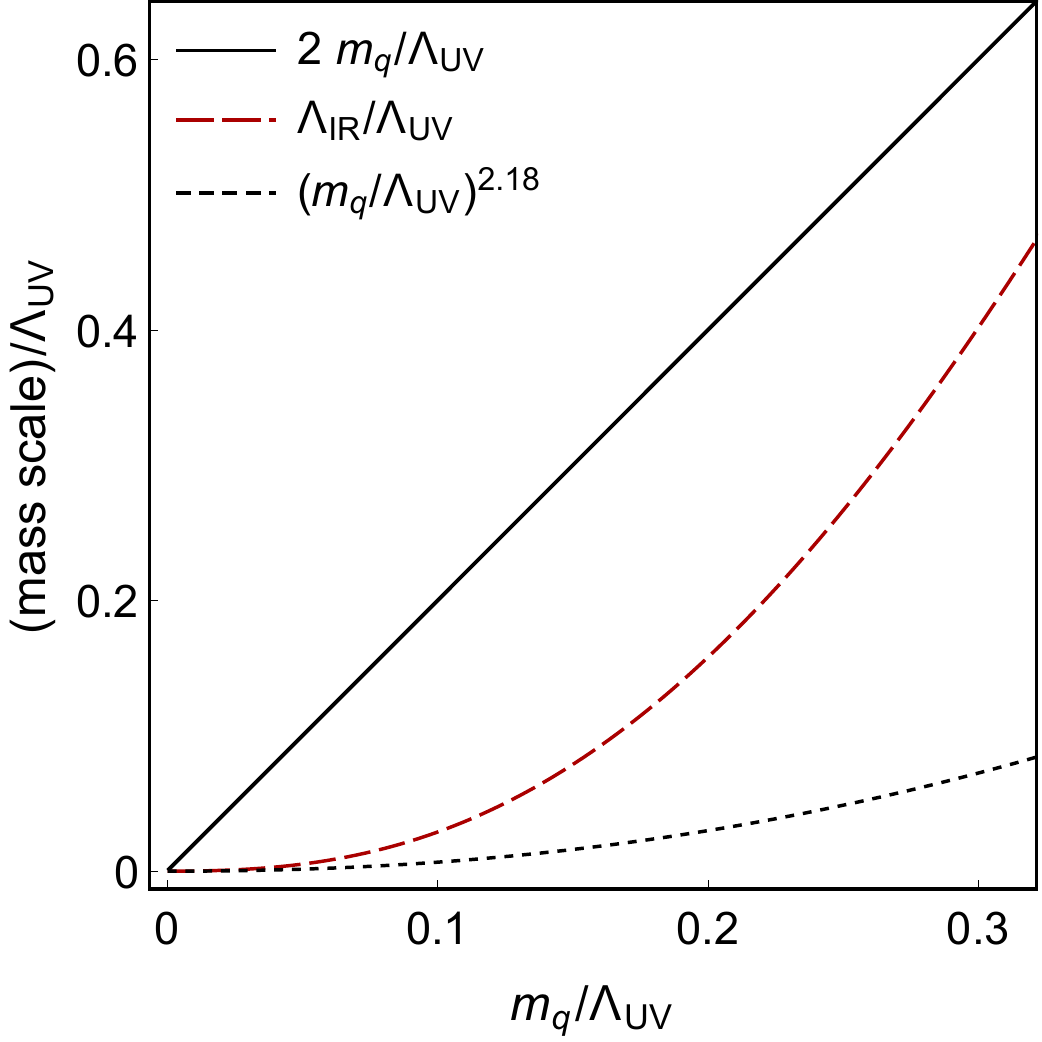}
\hfill
\includegraphics[height=0.35\linewidth,keepaspectratio]{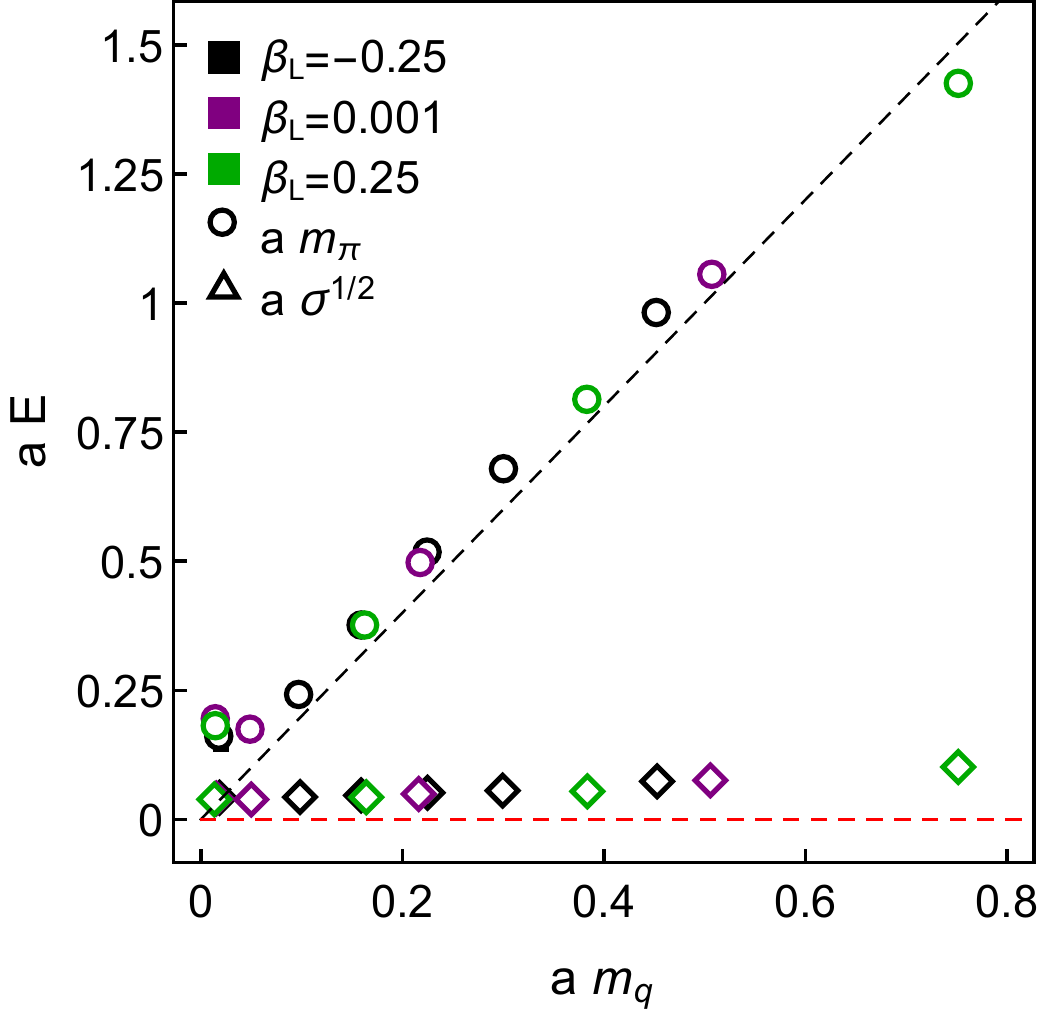}
\hfill~
\caption{{\em Left:} The expected hadron mass scale $2 m_q$ and the approximations of the confinement scale, Eq.~(\ref{approxLambda0}) and
evolution of Eq.~(\ref{eq:massivergeq}).  {\em Right:}
The pseudoscalar mass $m_\pi$ and the square root of string tension $\sigma^{1/2}$ as functions of $m_q$.  The data is measured from lattices of size $48^4$.}
\label{fig:scaling}
\end{figure}

On the other hand, because $m_q/\Lambda_\text{IR}$ grows as $m_q$ decreases, the 2-quark hadrons are effectively  ``heavy quark'' systems with masses
$m_{\text{Hadron}} \approx 2 m_q$.

A more accurate estimation of the confinement scale can be obtained with the massive $\beta$- and $\gamma$ -functions:
\begin{equation}
\frac{d g^2}{d \log\of{\lambda}} =-\beta(g^2,\lambda\,m)\ ,  ~~~~~~~
\frac{d \log\of{m}}{d \log\of{\lambda}}=\gamma(g^2,\lambda\,m)\ .
\label{eq:massivergeq}
\end{equation}
Here $\lambda$ is the length scale where coupling and mass are evaluated.
We use the background field momentum subtraction (BF-MOM) scheme at 2 loops \cite{Jegerlehner:1998zg,Dietrich:2010yw}, and
evolve the equations by setting $m_q$
at the initial scale $\lambda = 1/(2 m_{q,0})$ and evolve the equations to UV and IR.  The resulting $\Lambda_\text{IR}/\Lambda_\text{UV}$ is shown in Fig.~\ref{fig:scaling} as a function of $m_{q,0}/\Lambda_\text{UV}$.  We observe that this result agrees with the approximation \eqref{approxLambda0} at small $m_q$, but deviates from it substantially at larger $m_q$ when the mass threshold effects and higher order corrections affect the result significantly.

\begin{figure}[tbp]
\hfill
\begin{center}
\hfill
\includegraphics[width=0.35\linewidth,keepaspectratio]{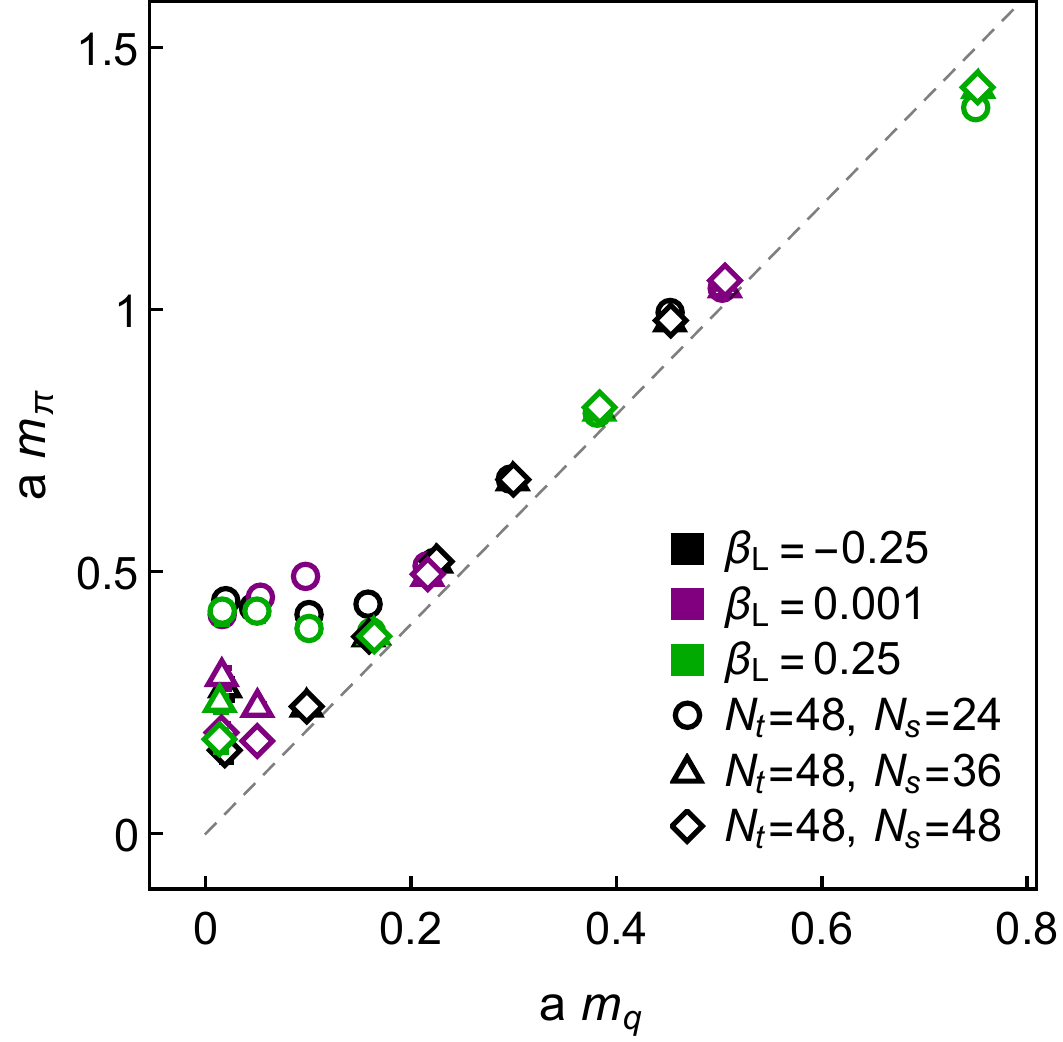}
\hfill
\includegraphics[width=0.35\linewidth,keepaspectratio]{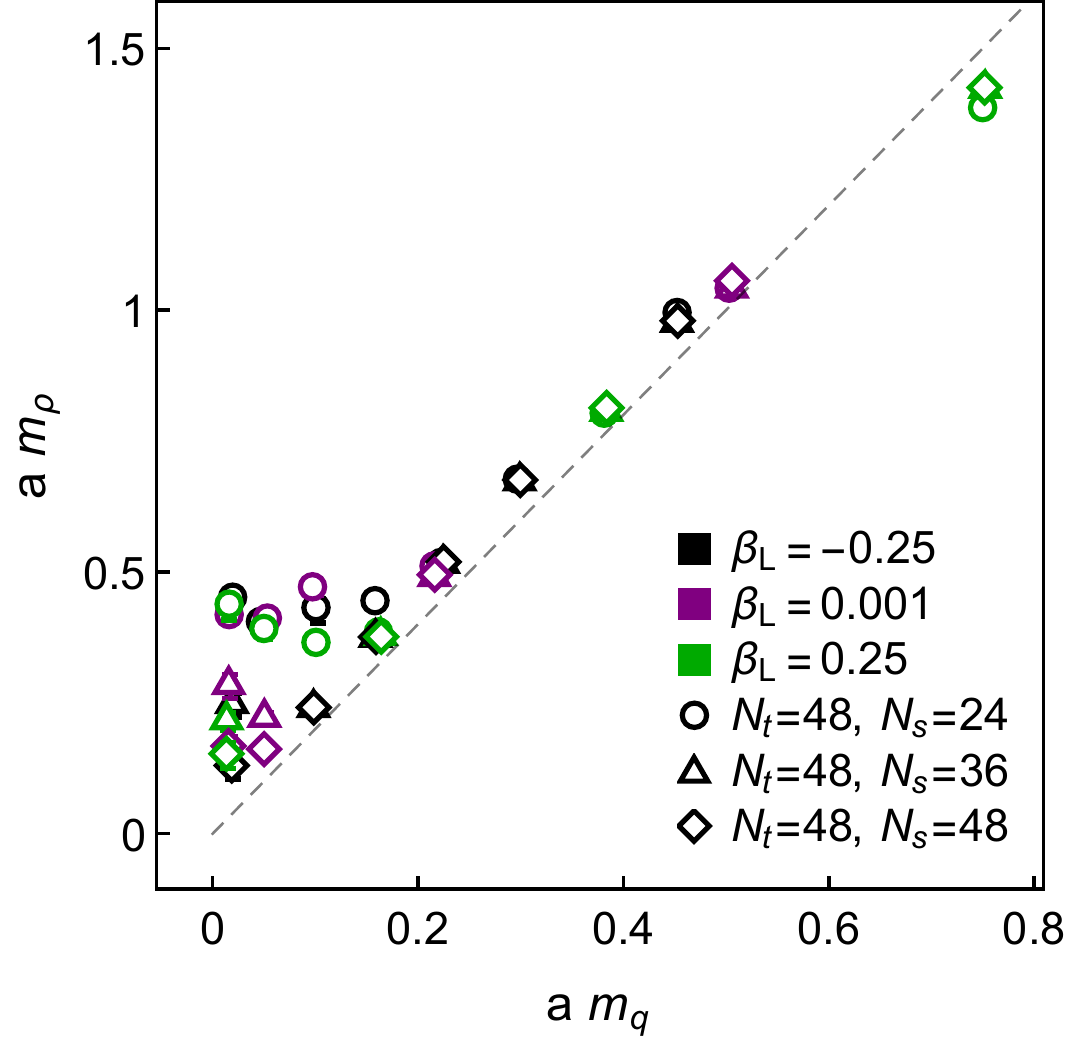}
\hfill~
\end{center}
\caption{The pseudoscalar meson mass $m_\pi$ (left) and the vector meson mass $m_\rho$ (right) as functions of the quark mass $m_{q}$ for $\beta_L =-0.25,\,0.001,\,0.25$.
The colours (black,purple,green) are used to distinguish between different values of $\beta$ and symbols (circle,triangle,diamond) are used to distinguish the different system sizes.
}
\label{fig:mpivsmq}
\end{figure}

In Fig.~\ref{fig:mpivsmq} we show the masses of pseudoscalar ($\pi$) and vector ($\rho$) meson, measured from different volumes and using three different bare inverse lattice couplings $\beta_L$.  The hadron masses are very close to the $2m_q$ -line at all $\beta_L$ and $\pi$ and $\rho$ are in practice degenerate.  There is a clear finite volume effect at small $m_q$, but the behaviour of the data makes it very plausible that the $m_{\rm Hadron} \approx 2m_q$ behaviour remains valid in the limit $m_q \rightarrow 0$ in infinite volume.

In order to characterise the confinement scale, we attempted to measure the string tension and scalar and tensor glueball masses with standard methods.
However, the confinement scale turns out to be very small, and we could not obtain meaningful results for the glueball masses and only an upper limit for the string tension $\sigma$ (for details, see ref.~\cite{Rantaharju:2015yva}).  On the right panel of Fig.~\ref{fig:scaling} we show the measurements of $a\,m_\pi$ and $a \sqrt{\sigma}$, and compare with the expected scaling on the left panel.  The meson masses obey the scaling well, and the string tension is also
It should be noted that on the left the quark masses are up to order $\Lambda_\text{UV}$, the Landau pole.  This domain cannot be reached in lattice simulations, and thus the lattice data corresponds to the left hand corner of the scaling plot.  We have arbitrarily related $a \Lambda_\text{UV} = 24$ here.

\section{Running coupling}

\begin{figure}[tbp]
\centering
\hfill
\includegraphics[width=0.35\linewidth,keepaspectratio]{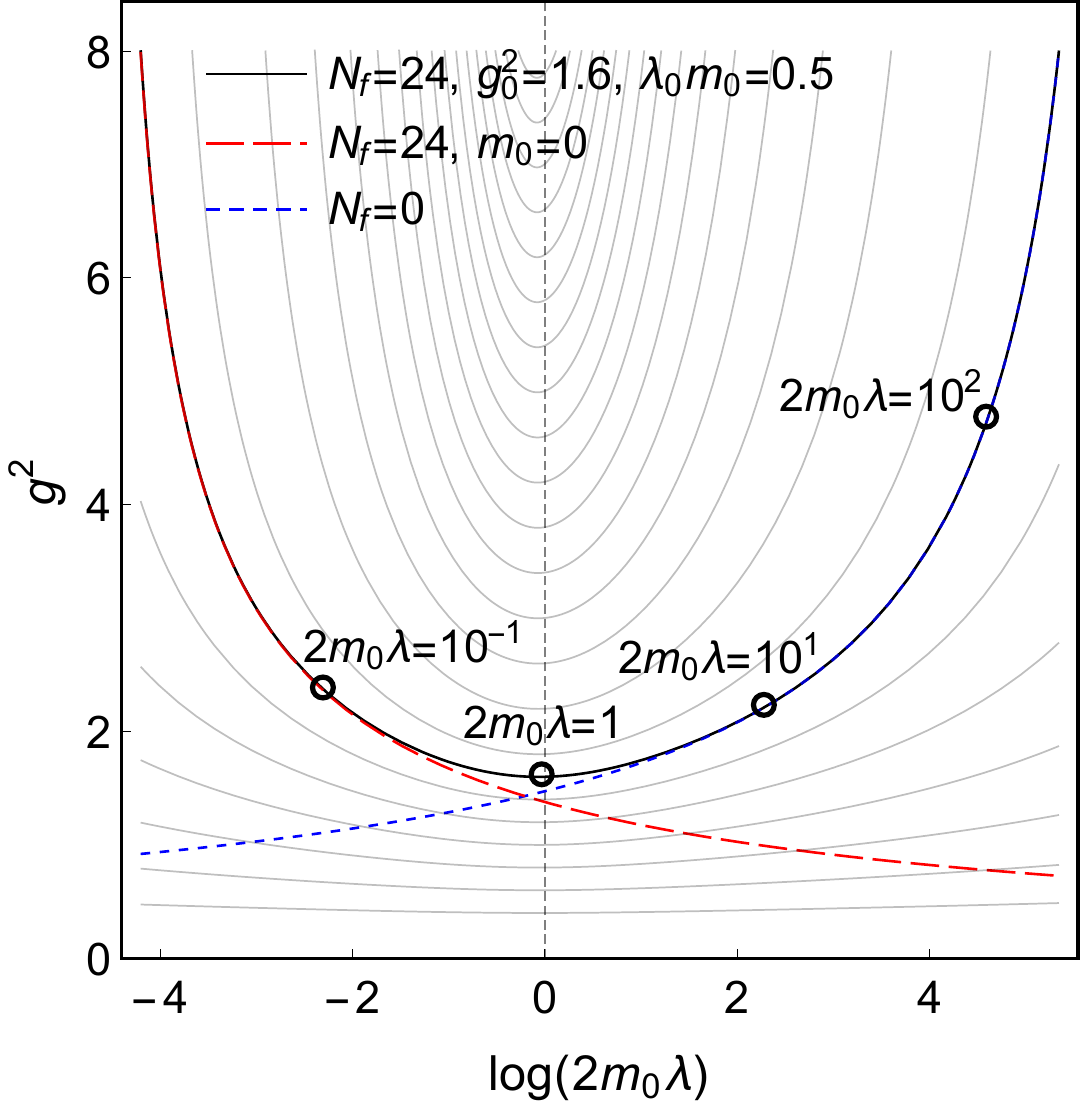}
\hfill
\includegraphics[width=0.35\linewidth,keepaspectratio]{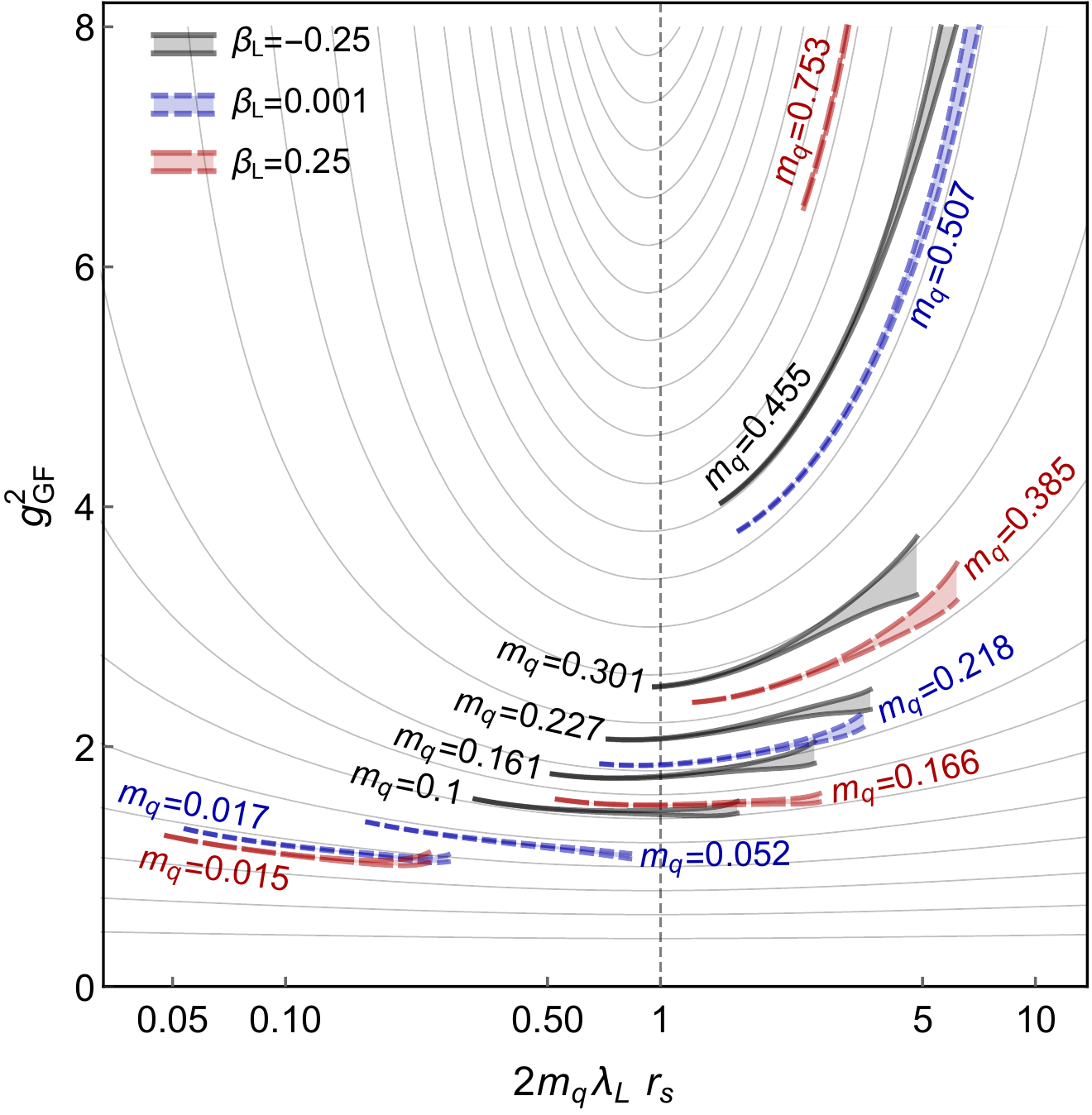}
\hfill~
\caption{{\em Left:}
The 2-loop perturbative evolution of the couplings $g^2$ as functions of the length scale $\lambda$, obtained by integrating Eqs.~(\ref{eq:massivergeq}) with varying initial value at $2m_0 \lambda = 1$.  To illustrate the asymptotic behaviour, pure gauge ($N_f=0$) and massless $N_f=24$ evolution have been matched with the massive evolution.  The small circles denote points where the coupling evolves when scale is changed by an order of magnitude.
{\em Right:}
The gradient flow data (black, blue and red error bands) as functions of $2\,m_q\,\lambda_L\,r_s$, where $m_q$ is the PCAC quark mass and $r_s=1/3$, superimposed on the perturbative curves.}
\label{fig:massiverunningcoupling}
\end{figure}

We obtain the 2-loop perturbative evolution of the coupling with the renormalization group equations (\ref{eq:massivergeq}).  The initial conditions are set by giving a set of values for $g^2(\lambda = 1/(2m_0))$.  The resulting curves are shown in Fig.~\ref{fig:massiverunningcoupling}.  It is evident that the coupling grows both at IR and UV ends, and that it asymptotically matches the pure gauge (IR) and massless $N_f=24$ (UV) behaviour.

The coupling is measured with the ``continuous $\beta$-function'' gradient flow method as described in section \ref{sec:lattice}.
The gradient flow coupling measurement is done with lattices of size $V=L^4$, $L=48a$, with smaller volumes used for finite size analysis.

\begin{figure}[tbp]
\centering
{\begin{minipage}[t]{0.33\linewidth}
\includegraphics[height=0.99\linewidth,keepaspectratio,margin=3pt 0pt,right]{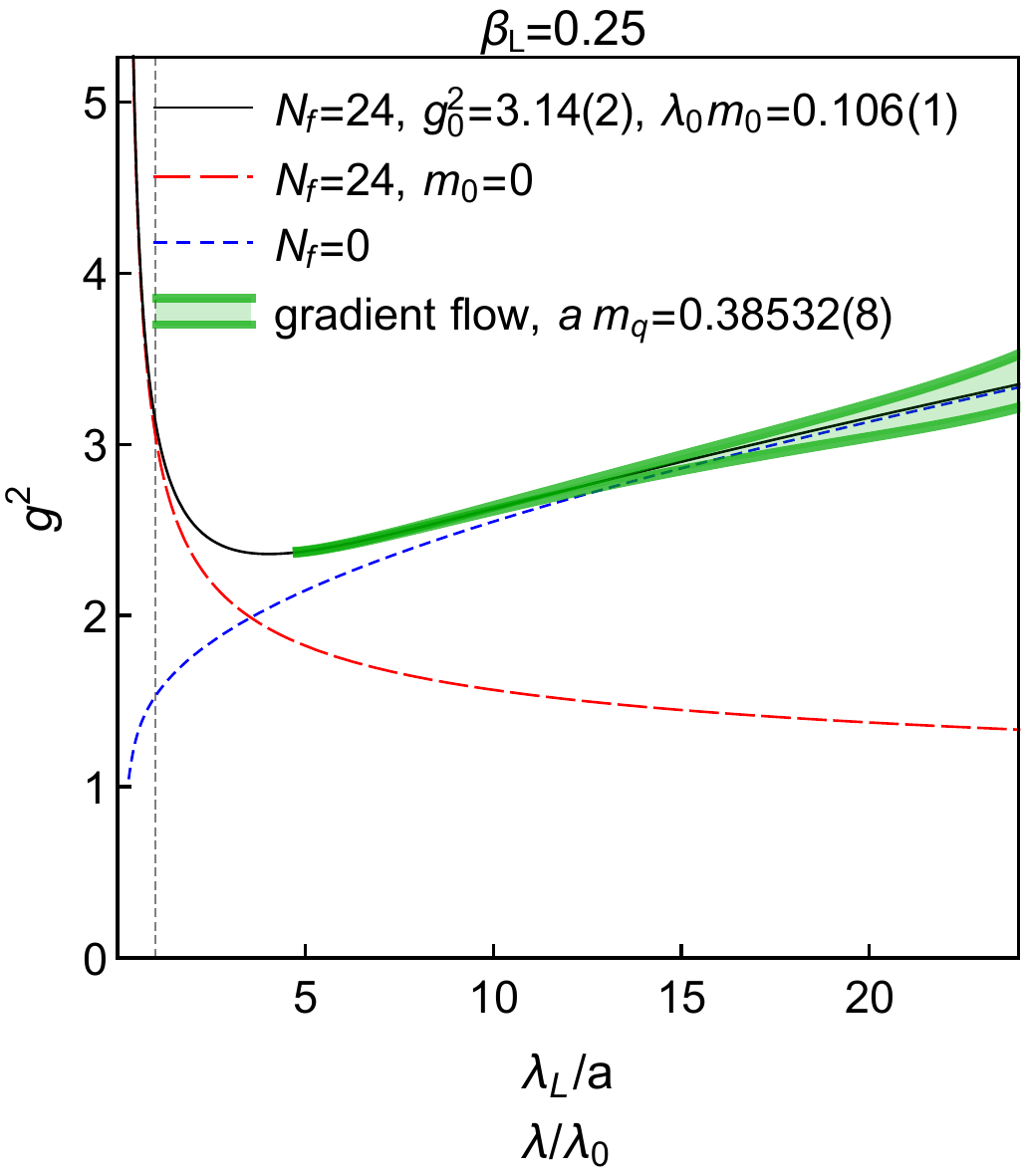}
\end{minipage}\hfill
\begin{minipage}[t]{0.33\linewidth}
\includegraphics[height=0.99\linewidth,keepaspectratio,margin=3pt 0pt,right]{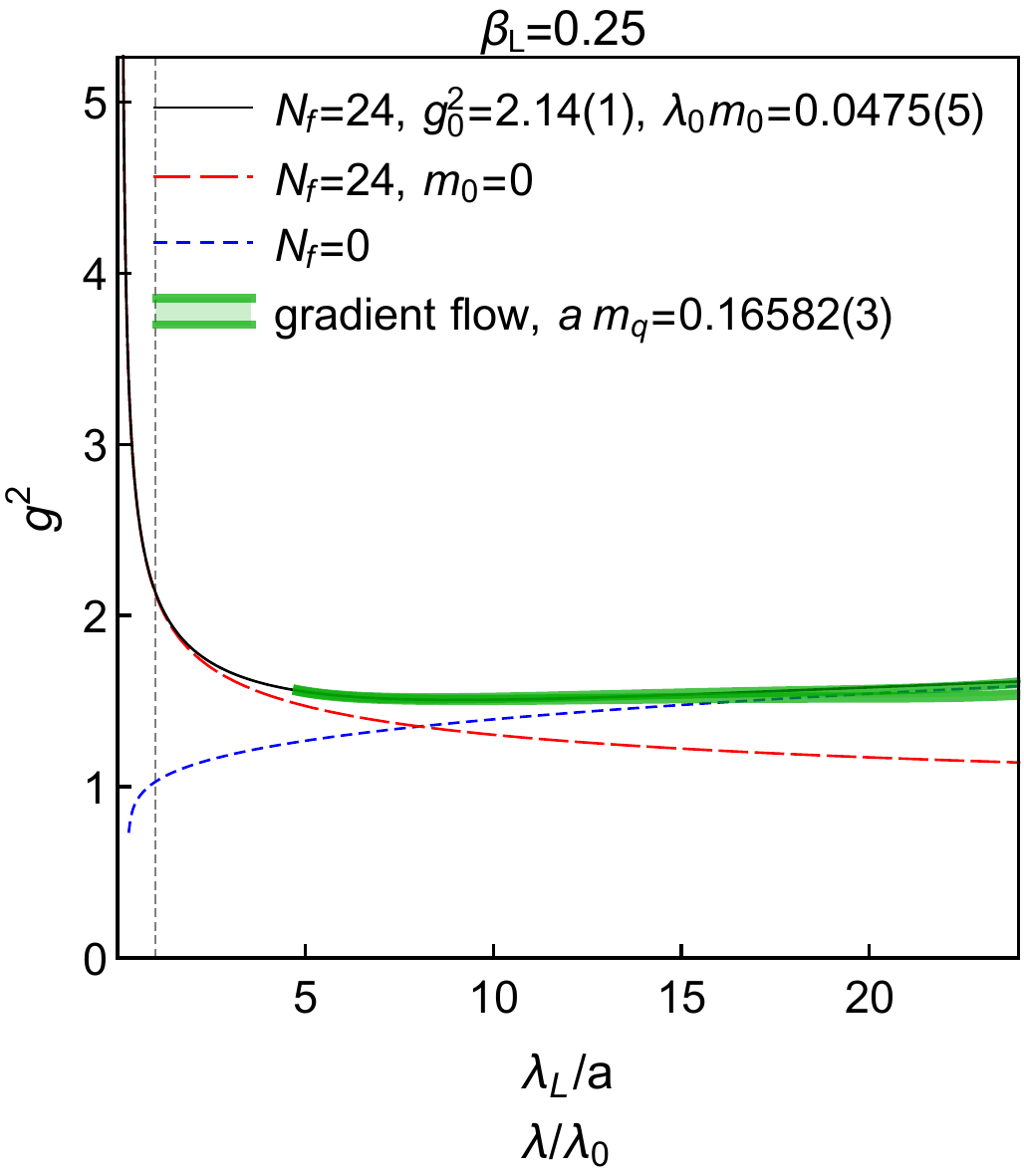}
\end{minipage}\hfill
\begin{minipage}[t]{0.33\linewidth}
\includegraphics[height=0.99\linewidth,keepaspectratio,margin=3pt 0pt,right]{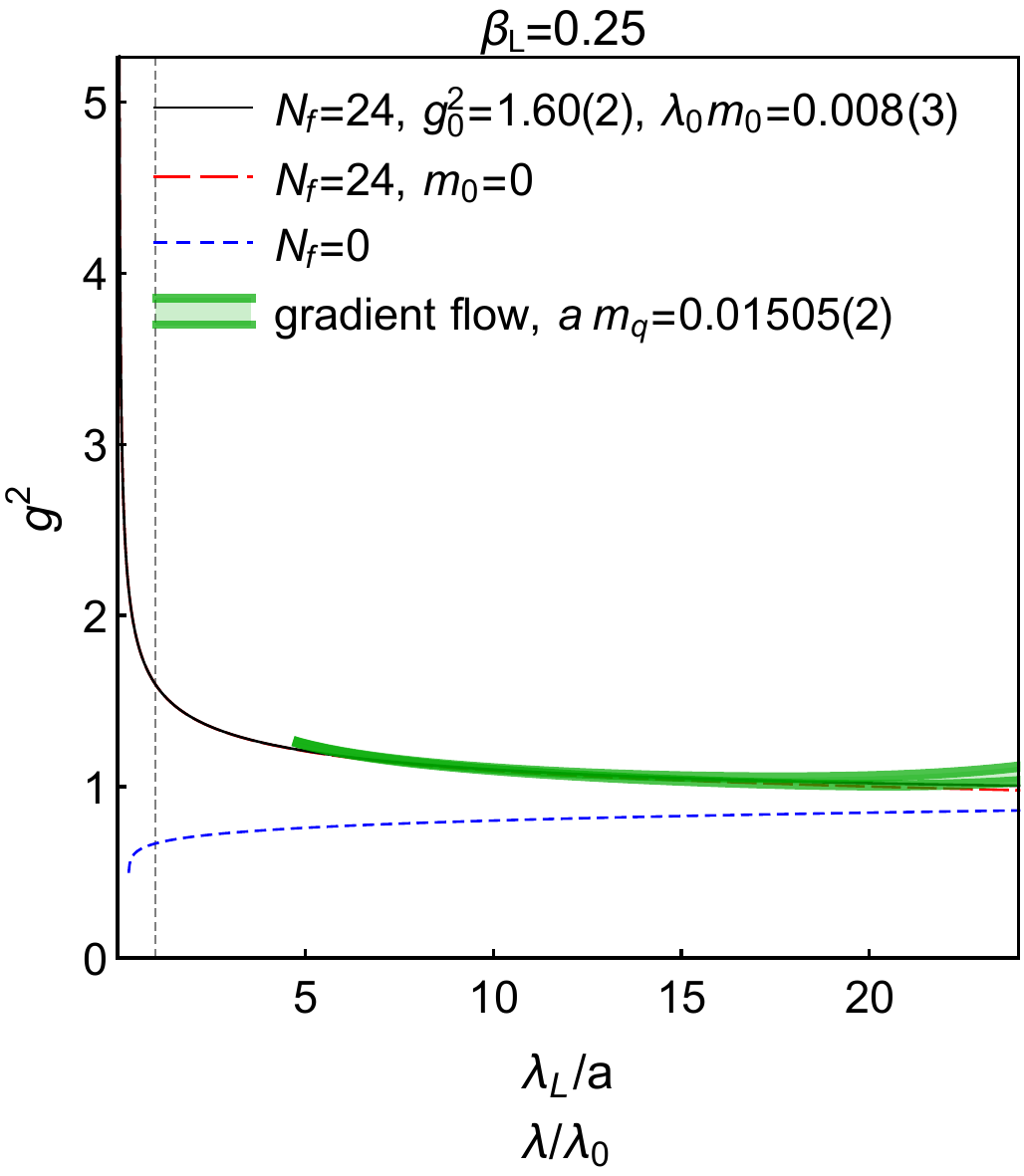}
\end{minipage}}\\[5pt]
\begin{minipage}[t]{0.33\linewidth}
\includegraphics[height=0.99\linewidth,keepaspectratio,margin=3pt 0pt,right]{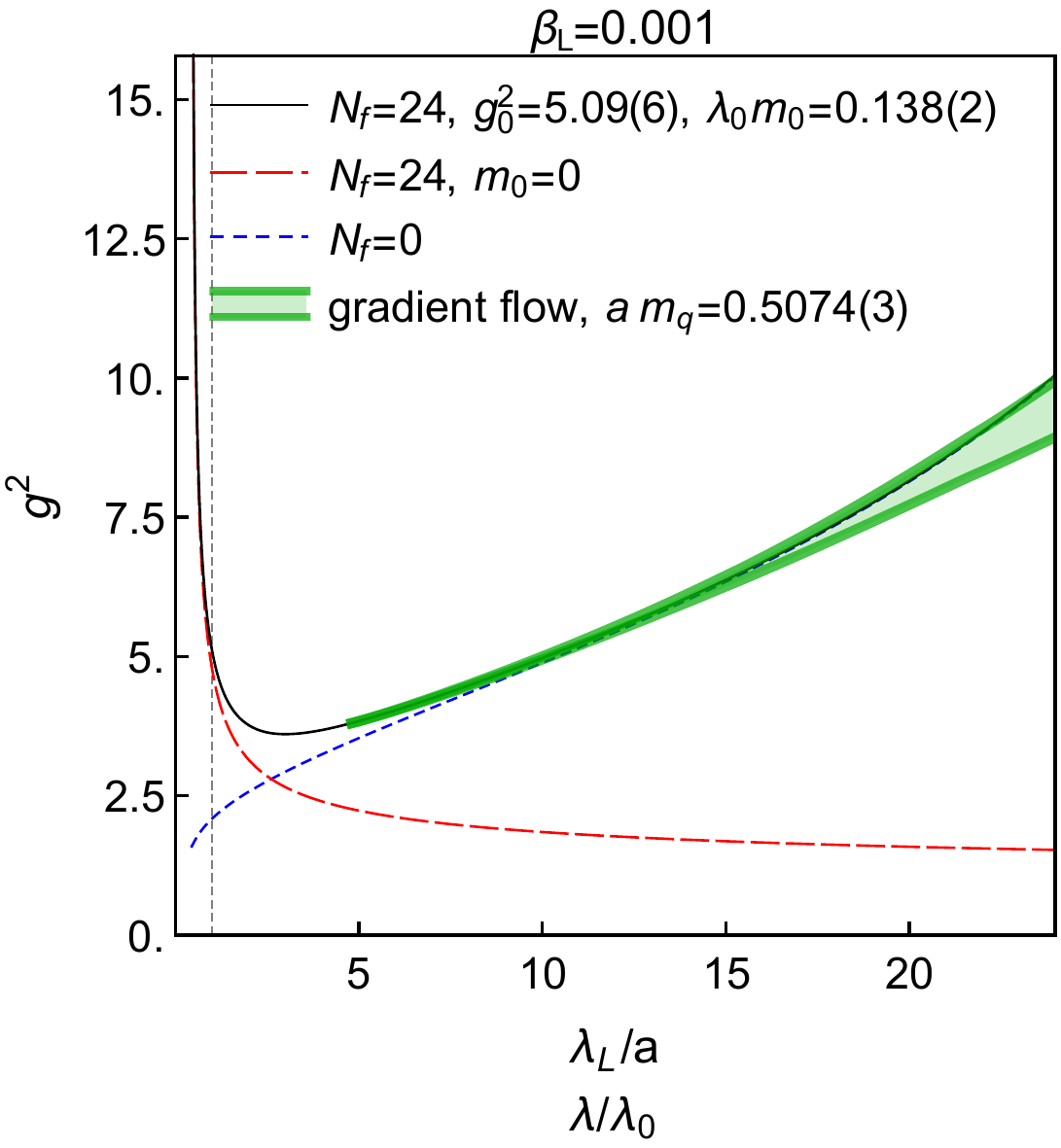}
\end{minipage}\hfill
\begin{minipage}[t]{0.33\linewidth}
\includegraphics[height=0.99\linewidth,keepaspectratio,margin=3pt 0pt,right]{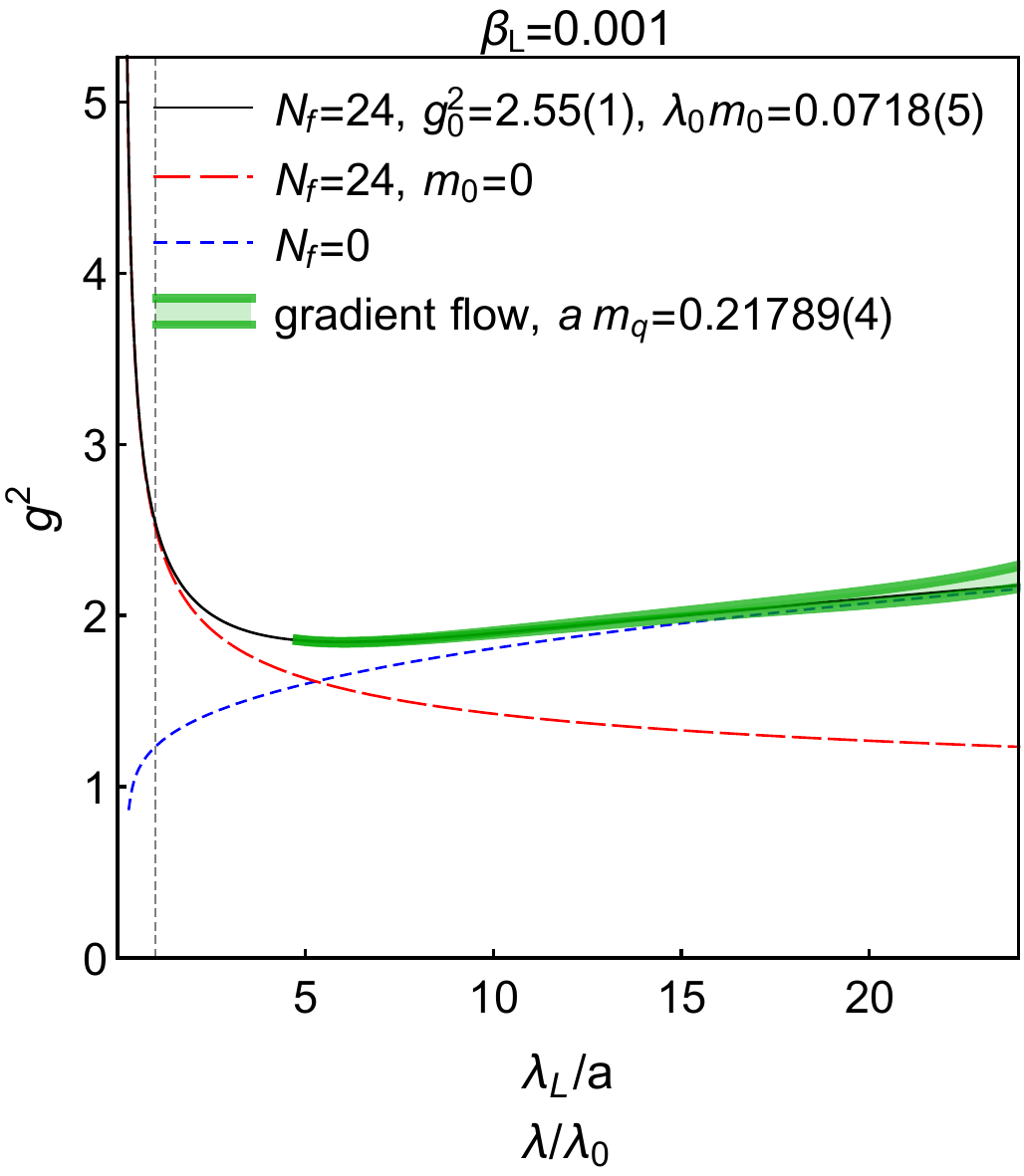}
\end{minipage}\hfill
\begin{minipage}[t]{0.33\linewidth}
\includegraphics[height=0.99\linewidth,keepaspectratio,margin=3pt 0pt,right]{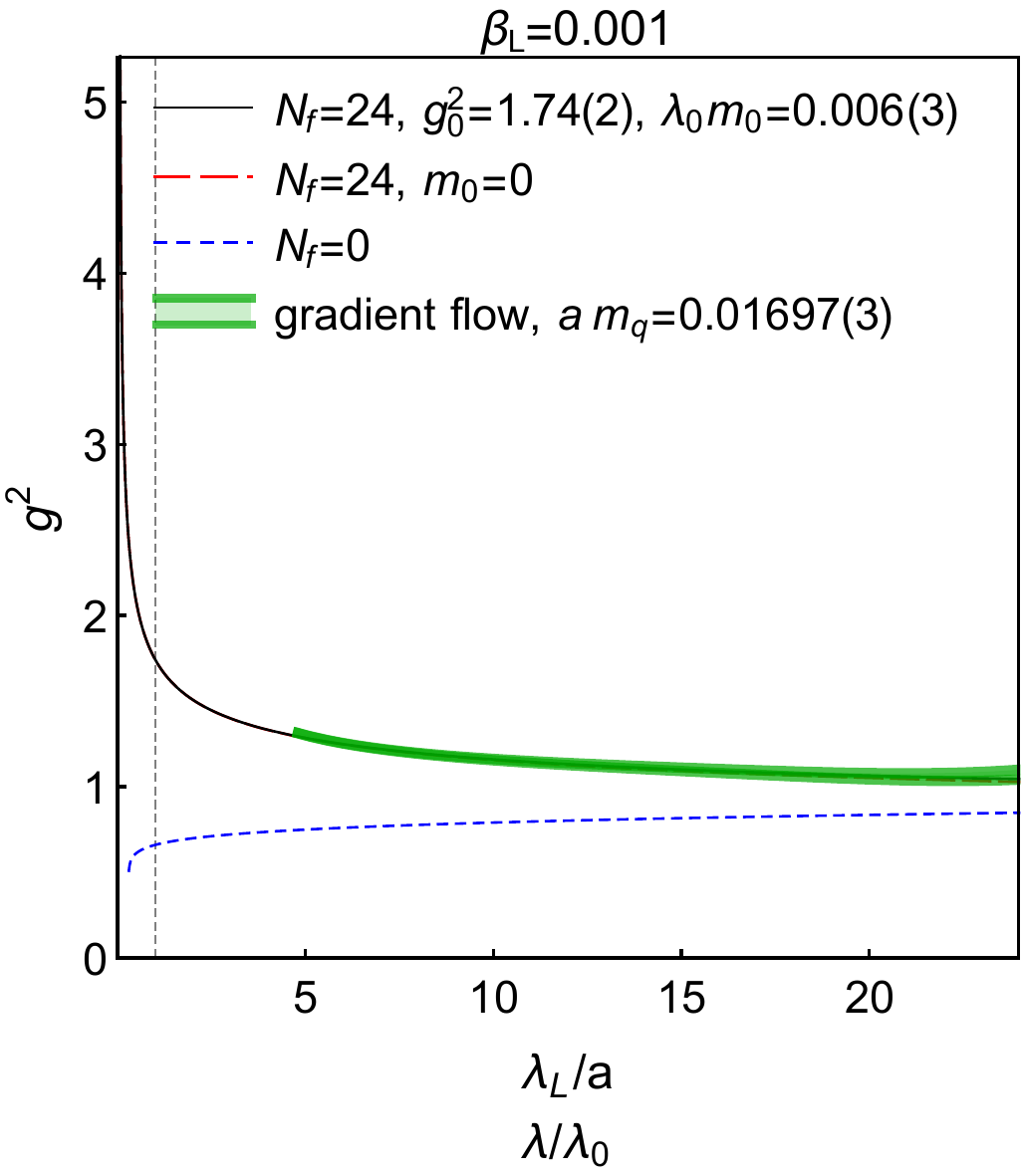}
\end{minipage}\\[5pt]
\caption{The measured gradient flow running couplings (green bands), obtained on a $V=(48a)^4$ lattice at $\beta\in\cof{0.25,0.001}$ at three quark masses $m_q$, to which two-loop running coupling is fitted (solid black line). The gradient flow length scale $\lambda_L$ is shown in interval $\lambda_L/a \in [4.8,24]$.
In comparison, the matched pure gauge SU(2) (blue dotted line) and $N_f=24$ $m_q=0$ (red dashed line) couplings are also shown.
}
\label{fig:couplingvsscalefit}
\end{figure}

Fig.~\ref{fig:couplingvsscalefit} includes examples of $\gGF(\lambda_L)$ at tree different values of $m_q$ for $\beta_L \in {0.25,0.01}$. The switch from the ``heavy quark'' (left column) to the ``light quark'' behaviour (right column) is evident also on the lattice.
We compare the results with the 2-loop perturbation theory by fitting $g_0^2$ and $\lambda_0 m_0$ to match $\gGF$. In effect, the fit procedure achieves the relative multiplicative renormalization between $\lambda m_0$ in BF-MOM scheme and $\lambda_L m_q$ as measured from the lattice.

It turns out that the relative renormalization of $\lambda m$ between lattice and BF-MOM schemas is roughly constant in our range of masses.
Indeed, in Fig.~\ref{fig:couplingvsscalefit} we plot all measurements of $\gGF$ against $2m_q\lambda_L/3$, overlaid with the perturbative $g^2$ from Fig.~\ref{fig:massiverunningcoupling}.  Besides the factor of $1/3$ there are no fitted parameters.  The lattice data follows the 2-loop perturbative curves remarkably well, independent of the value of $\beta_L$.
There are cases where simulation results with different $\beta_L$ and $m_q$ fall on curves which are very close to each other. Since different values of $\beta_L$ correspond to different lattice spacings, this demonstrates that the results scale when lattice spacing is varied. In contrast to the asymptotically free lattice QCD, the lattice spacing becomes smaller when $\beta_L$ is decreased, and the theory does not have a continuum limit because of the UV Landau pole.  For details we refer to ref.~\cite{Rindlisbacher:2021hhh}.

\section{Conclusions}

SU(2) gauge theory with $N_f=24$ quarks provides an interesting test case for studying decoupling of quarks.
At energy scales $\mu \gg m$ the coupling grows with the energy, reaching the UV Landau pole, whereas at $\mu \ll m$ the system behaves as pure gauge SU(2) theory where coupling decreases when energy grows.  We demonstrate clear non-perturbative evidence of this behaviour and the quark decoupling at energy scale $\mu \sim m$.
These features have consequences for the physical excitation spectrum as the quark mass is varied, and we have presented scaling laws describing the hadron mass and the confinement scale behaviour as functions of the quark mass.  These results provide a consistent non-perturbative description of the behaviour of the theory from IR to UV scales.

\subsubsection*{Acknowledgements}
We acknowledge the support of the Academy of Finland grants 308791, 310130 and 320123, and CSC -- IT Center for Science, Finland, for generous computational resources.

\end{document}